\documentclass[11pt,a4paper]{article}
\usepackage{amsmath,epsfig,indentfirst,amssymb,graphicx}

\begin{document}

\begin{center}

{\LARGE {\bf Evaluation of the Primary Energy of UHE Photon-induced
Atmospheric Showers from Ground Array Measurements}}

\vspace{1cm}

   Pierre Billoir, C\'ecile Roucelle, Jean-Christophe Hamilton\\
   LPNHE Paris, Univ. Paris VI-VII \& CNRS

\end{center}
\vspace{5mm}

\abstract{

A photon induced shower at $E_{prim}\ge 10^{18}$ eV exhibits very specific features and is different from a hadronic one. At such energies, the LPM effect (\cite{LP}, \cite{M}) delays in average the first interactions of the photon in the atmosphere and hence slows down the whole shower development. They also have a smaller muonic content than hadronic ones. The response of a surface detector such as that of the Auger Observatory  to these specific showers is thus different and has to be accounted for in order to enable potential photon candidates reconstruction correctly. The energy reconstruction in particular has to be adapted to the late development of photon showers. We propose in this article a method for the reconstruction of the energy of photon showers with a surface detector. The key feature of this method is to rely explicitly on the development stage of the shower. This approach leads to very satisfactory results ($\simeq 20\%$). At even higher energies ($5.10^{19}$ eV and above) the probability for the photon to convert into a pair of e$^+$e$^-$ in the geomagnetic field becomes non negligible and requires a different function to evaluate the energy with the proposed method. We propose several approaches to deal with this issue in the scope of the establishment of an upper bound on the photon fraction in UHECR.

}

\section{General framework}
  This study is aimed to analyze the response of a surface detector (SD) of
extensive atmospheric showers induced by primary photons of ultra high
energy (more than 10$^{18}$ eV). Particular applications will be made for the
Auger Observatory, where the Surface Detector is an array of water Cherenkov
tanks, at an average altitude of 1400 m a.s.l. \cite{NIM2004}.

\section{Energy reconstruction of extensive air showers with the Auger surface detector using a Monte Carlo based calibration}

Measuring the primary energy $E_{prim}$ of an extensive atmospheric shower
from measurements in a sparse ground array is not straightforward. Classical
evaluations for showers induced by protons or nuclei use
a relation between $E_{prim}$ and the signal interpolated at a given distance
$r_0$ from the shower axis (\cite{2002-075})~:
$$ S(r_0)= E_{prim}^{\alpha}\,f(\theta)~~~~~
(\theta :{\rm zenith}~{\rm angle}) $$
The exponent $\alpha$ is slightly less than 1 (typically 0.95), to account for
the longitudinal stretching of the shower, increasing with $E_{prim}$.
The function $f(\theta)$ includes essentially the following ingredients:
\begin{itemize}
\item the longitudinal evolution of the shower: the size at ground level
depends on the slant depth $X=X_{ground}/\cos\theta$, where $X_{ground}$ is
the  vertical thickness at ground level (practically constant for a given
detector site).
\item the dependence of the signal to the incidence angle of the shower
particles, determined by the type and the geometry of the detector.
\item the sensitivity of the detector to different particles: a thin
scintillator counts the charged particles (mainly electrons and positrons
for moderately inclined showers), while a Cherenkov water tank sees the photon
by their internal cascading, and has an enhanced
sensitivity to muons: $f(\theta)$ will be more sensitive to the muonic profile,
which is different from the electromagnetic one.
\end{itemize}
In usual conditions, for a detector at moderate altitude, the maximum $X_{max}$
of the longitudinal profile is above or around  the ground level
($X_{max} \lesssim X_{ground}$): inclined showers hit the ground in their
decreasing phase; moreover, the acceptance does not increase with $\theta$,
so that $f(\theta)$ is a decreasing function. 

\par
\begin{figure}[!h]
\begin{center}
\includegraphics[scale=0.4]{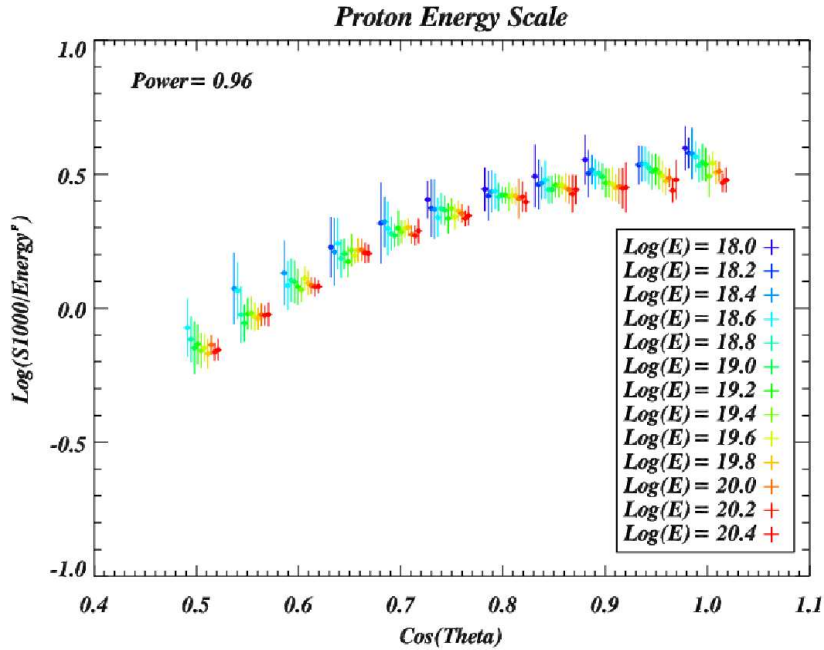}
\caption{\small{Relation between primary energy and signal at 1000 m
from the core, for simulated proton showers (AIRES), at various zenith angles and
energies. Energies are taken in EeV and $S(1000)$ is given in VEM. The exponent of the energy P is there set at 0.96, as indicated. For a given value of $\cos\theta$, the points are shifted to make the plot more readable.}}
\label{proton}
\end{center}
\end{figure}

 To test these features, we used samples of showers generated with AIRES~\cite{aires}
from both protons and photons, in the energy range 10$^{18}$ to 10$^{20.5}$ eV,
and we apply a standard detector simulation for the Auger array, and an event
reconstruction procedure to evaluate the signal in each tank and the
interpolated signal $S(1000)$. As can be seen for protons simulations on fig.~\ref{proton},
the factorization is approximately valid for showers initiated by protons or nuclei, because
the shower-to-shower fluctuations of the longitudinal profile and of the
muon/electromagnetic ratio are not too weak. A bias is expected as a
function of the primary mass (mass number $A$) and the primary energy, but
studies on simulated events suggests than the above formula can give a
precision of the order of 20 \% on $E_{prim}$ if $S(r_0)$ is precisely
known, and if, of course, modelling errors in shower simulations may be
neglected (see~\cite{2002-075}). 
\\

\par
\begin{figure}[!h]
\begin{center}
\includegraphics[scale=0.8]{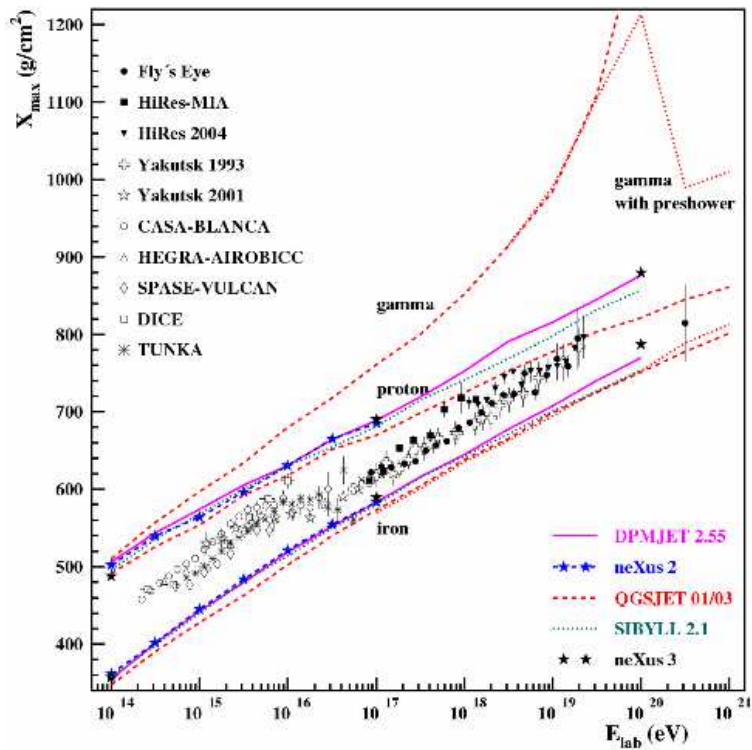}
\caption{\small{Average depth of shower maximum $X_{max}$ versus energy simulated for pri- 
mary photons, protons and iron nuclei. Depending on the specific particle trajectory 
through the geomagnetic field, photons above $\simeq 5.10^{19}$eV can create a preshower: 
as indicated by the splitting of the photon line, the average $X_{max}$ values then do not 
only depend on primary energy but also arrival direction. For nuclear primaries, cal- 
culations for different hadronic interaction models are displayed (QGSJET01~\cite{qgsjet1}, 
QGSJETII~\cite{qgsjet2}, SIBYLL2.1~\cite{sibyll}). Also shown are experimental data (for references 
to the experiments, see~\cite{Autres}).}}
\label{elong}
\end{center}
\end{figure}

\begin{figure}[!h]
\begin{center}
\includegraphics[scale=0.4]{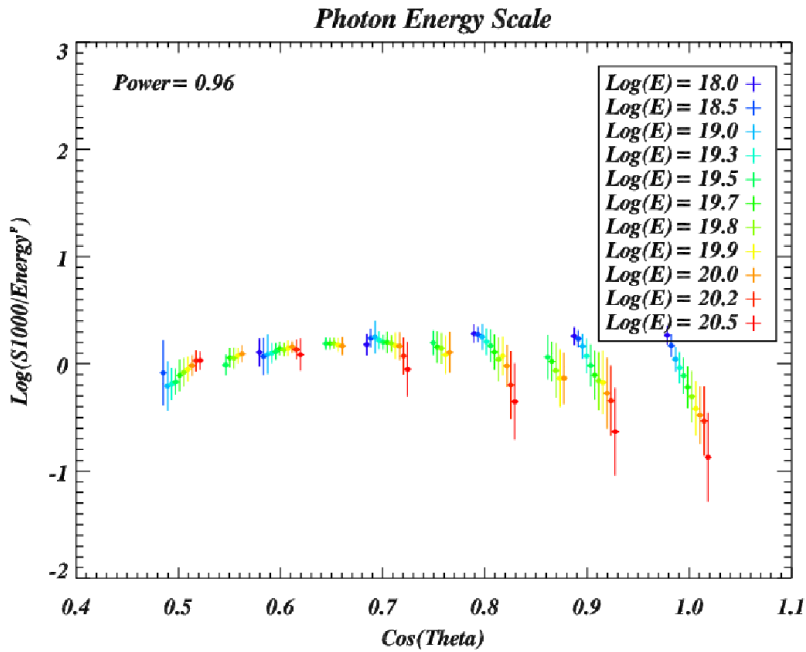}
\caption{\small{Relation between primary energy and signal at 1000 m
from the core, for photon induced showers (without pre-showering in the
geomagnetic field), with the same conventions as in fig.~\ref{proton}.}}
\label{gamma}
\end{center}
\end{figure}

 The situation is expected to be quite different for photon induced
showers. First, their muonic content is very low compared to a shower initiated by a proton or a nucleus. Secondly, their longitudinal
development is slower, especially for $E_{prim} \gtrsim 10^{19} eV$, where
the LPM effect delays the first steps of the electromagnetic cascade.
Simulations indicate that the average value of $X_{max}$ exceeds 900 g/cm$^2$
at $E_{prim}=10^{19}$ eV, and increases rapidly with $E_{prim}$ (see fig.~\ref{elong}). Then, nearly
vertical showers reach the ground before their maximum, and for a given primary
energy, inclined showers may have a larger density than vertical ones. The
factorization is no more valid, and $f(\theta)$ should be replaced by
a function strongly dependent of $E_{prim}$. This is clear on Fig.~\ref{gamma},
for the same energy range as in Fig.~\ref{proton}.

\par Moreover, above a threshold energy depending on the strength of the
geomagnetic field (5.10$^{19}$ eV for the southern site of the Auger
Observatory, \cite{1998-049}), the photon may undergo a conversion into an $e^+e^-$ pair with
a non-negligible probability before entering the atmosphere \cite{MacBreen}~; the electrons
themselves radiate hard synchrotron photons: this causes a electromagnetic
``pre-showering'' and the particles entering the atmosphere are mainly below
the LPM threshold \cite{CorsConv}, \cite{1998-049}. Then the atmospheric shower will have a much faster
development than the one induced by an unconverted photon at the same primary
energy (similar to showers from a photon in the range 10$^{18}-$10$^{19}$ eV). This point will be developped in section \ref{conv}.

\section{The ``universal profile'' picture}
\label{univ}
\par This picture is based on the fact that the ``remote electromagnetic
profile'' (evolution of the density of photons, electrons, positrons with
the depth, at a large distance from the core) follows closely the global
profile (total number of charged particles in the shower), with a delay due
to the the lateral diffusion (about 150 g/cm$^2$ at 1000 m from the axis).
If we neglect the contribution of the muons to the signals, and if we suppose
that the detector response varies smoothly with the incidence angle, we expect
that the ratio $S(r_0)/E_{prim}$, when expressed as a function of 
$X=X_{ground}/\cos\theta$, will behave as a delayed profile, which, in first
approximation, is an universal function of $X-X_{max}$, with a shape similar to
the Gaisser-Hillas function commonly used to describe the global profile.
\par
\begin{figure}[!h]
\begin{center}
\includegraphics[scale=0.5]{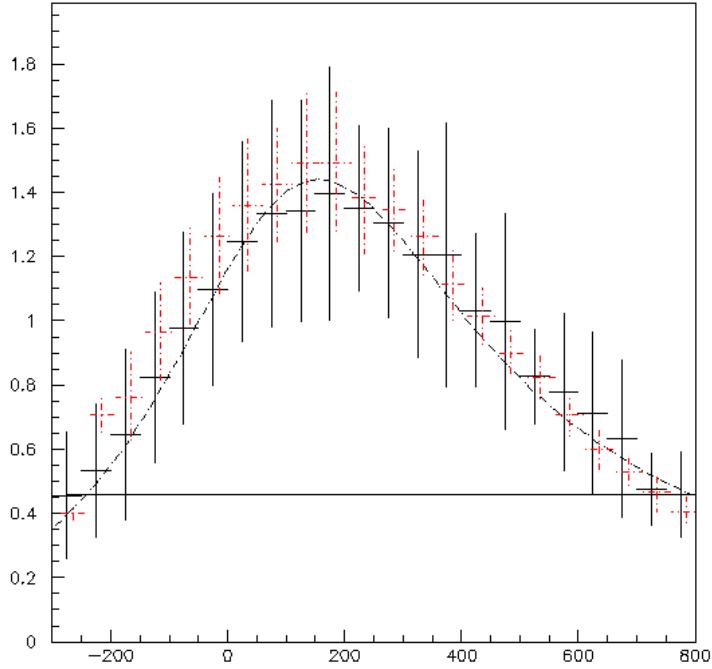}
\caption{\small{Ratio $S(1000)/E_{prim}$ as a function of $X_{ground}/\cos\theta-X_{max}$
for photon induced showers. Solid (black): unconverted photons; dash-dotted
(red): magnetically converted photons. The curve is a fit on the unconverted
ones.}}
\label{univ_plot}
\end{center}
\end{figure}

 Actually, plotting the ratio $S(1000)/E_{prim}$ as a function of
$\Delta X = X_{ground}/\cos\theta-X_{max}$ for simulated photon showers
at various energies and zenith angles (see Fig. \ref{univ_plot}), we obtain a
a overall curve with a maximum around 150 g/cm$^2$.

It is remarkable that showers from magnetically converted photons follow a very
similar law, in spite of the large differences in $X_{max}$; moreover, their
dispersion is reduced, because they give a superposition of subshowers at lower
energies. The observed profile is not
exactly represented by a Gaisser-Hillas function, because of the small muonic
fraction, the detector shape effects, and the fact that the end of the descent
is less steep at large distance than in the global profile. We use a empirical
parametrization:
$$ S(1000)/E_{prim} = 1.4~\frac{1+\frac{\Delta X-100}{1000}}
  {1+\left(\frac{\Delta X-100}{340}\right)^2}~~{\rm VEM/EeV} $$
where the depths are in g/cm$^2$; by convention VEM represents the signal given
by a vertical muon.
\par
\begin{figure}[!h]
\begin{center}
\includegraphics[scale=0.5]{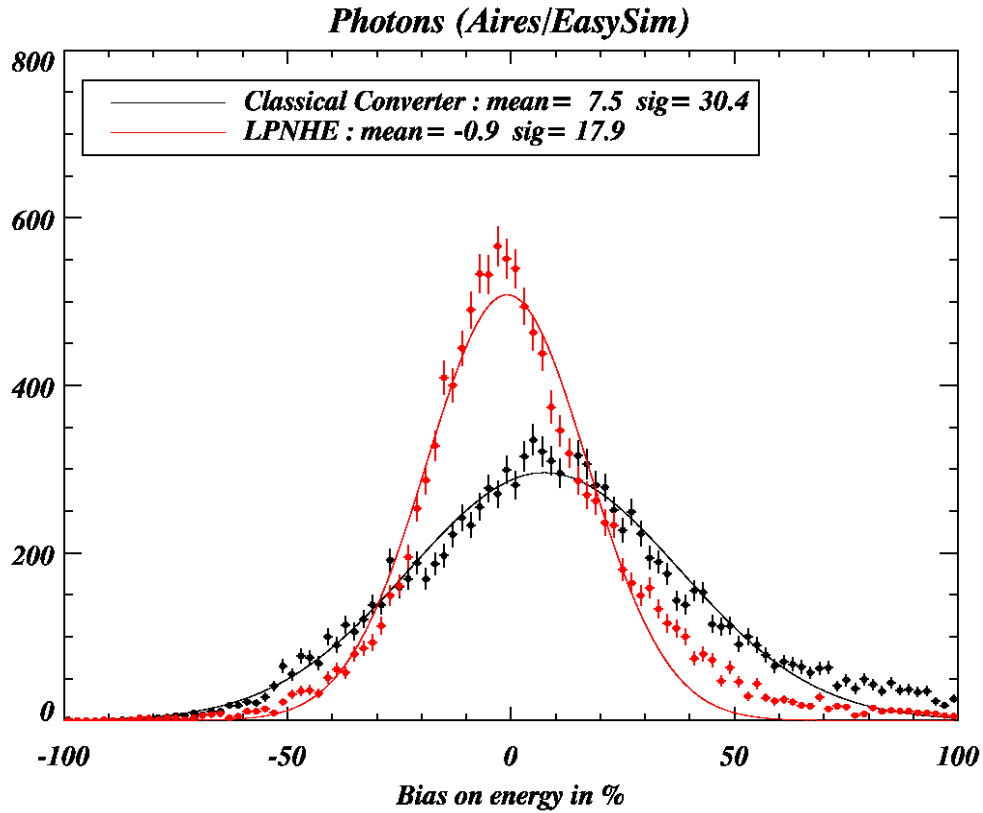}
\caption{\small{Resolution on photon energy. Black: ``proton-like'' method
(factorization); red: this method. 
zenith angles are taken between 35$^{\circ}$ and 60$^{\circ}$. Simulated energies from $10^{19}$ eV to $10^{19.4}$ eV have been taken there.}}
\label{resol}
\end{center}
\end{figure}

 To fully exploit this relation we have to know the value of $X_{max}$;
for events seen by the ground detector only, $X_{max}$ is not measured, then
we can use an average dependence on energy, deduced from the simulations (for
photons without pre-showering):
$$ X_{max}=856+141\,log_{10}(E_{prim})~~~~~{\rm with}~~E_{prim}~{\rm in~EeV} $$
If we suppose here that both $S(1000)$ and $\theta$ (hence $X$) are reliably
measured, even if the primary is a photon, then an iterative procedure may be
applied, starting with a rough estimation of the energy (e.g. twice the value computed
in the proton hypothesis, as we are expecting a clear underestimation of a factor 2 to 4 of the energy as we use the classic reconstruction)~: 
\begin{itemize}
\item estimate $X_{max}$ from the above formula
\item estimate $E$ from $S(1000)$ and $X-X_{max}$
\end{itemize}
In most cases the convergence is fast; however, if the value of $X-X_{max}$ is
in the beginning of the ascending phase of the profile, the iteration may
be problematic: a small value of $S(1000)/E$ gives a large value of $E$,
possibly larger than the previous estimation, pushing down $X-X_{max}$,
then one finds a smaller $S(1000)/E$, and so on. If this positive feedback
is large, the fluctuations of $X_{max}$ produce large fluctuations on $E$,
and sometimes the iteration diverges. On the contrary, in the descending
phase, the feedback is negative.
  For these reasons, we discard the left tail on the profile by setting a
minimum value of -50 g/cm$^2$ to $X-X_{max}$. As a consequence, we may
underestimate the energy of nearly vertical showers with a late development.
Fig.~\ref{resol} shows that the resolution on energy can be significantly improved,
compared to a ``proton-like'' evaluation with a simple power law. The resolution achieved with this method is roughly 20\% even up to $10^{20}$ eV. Nevertheless, distribution tails shown here would degrade gradualy at higher energies because of shower to shower important fluctuations, as discussed just below.  
It may be safe to put a lower cut on the zenith angle (at $\simeq 35^{\circ}$ for example) to avoid the problem of showers seen well before their maximum (let us note that anyway the trigger
acceptance us suppressed for such showers).  

\section{Handling magnetically converted photons}
\label{conv}

\begin{figure}[!h]
\begin{center}
\resizebox{10cm}{!}{\centering{\includegraphics{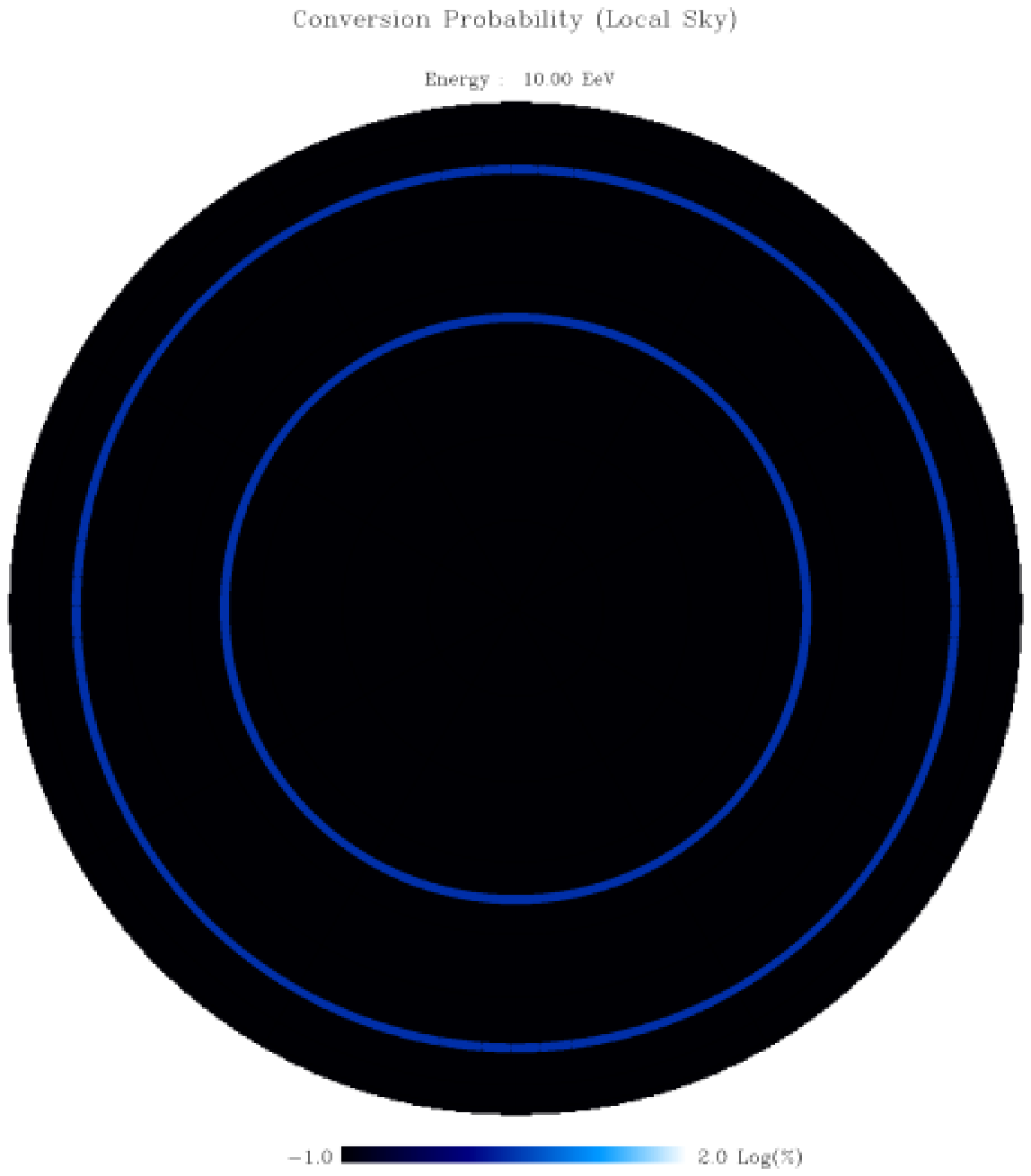}\includegraphics{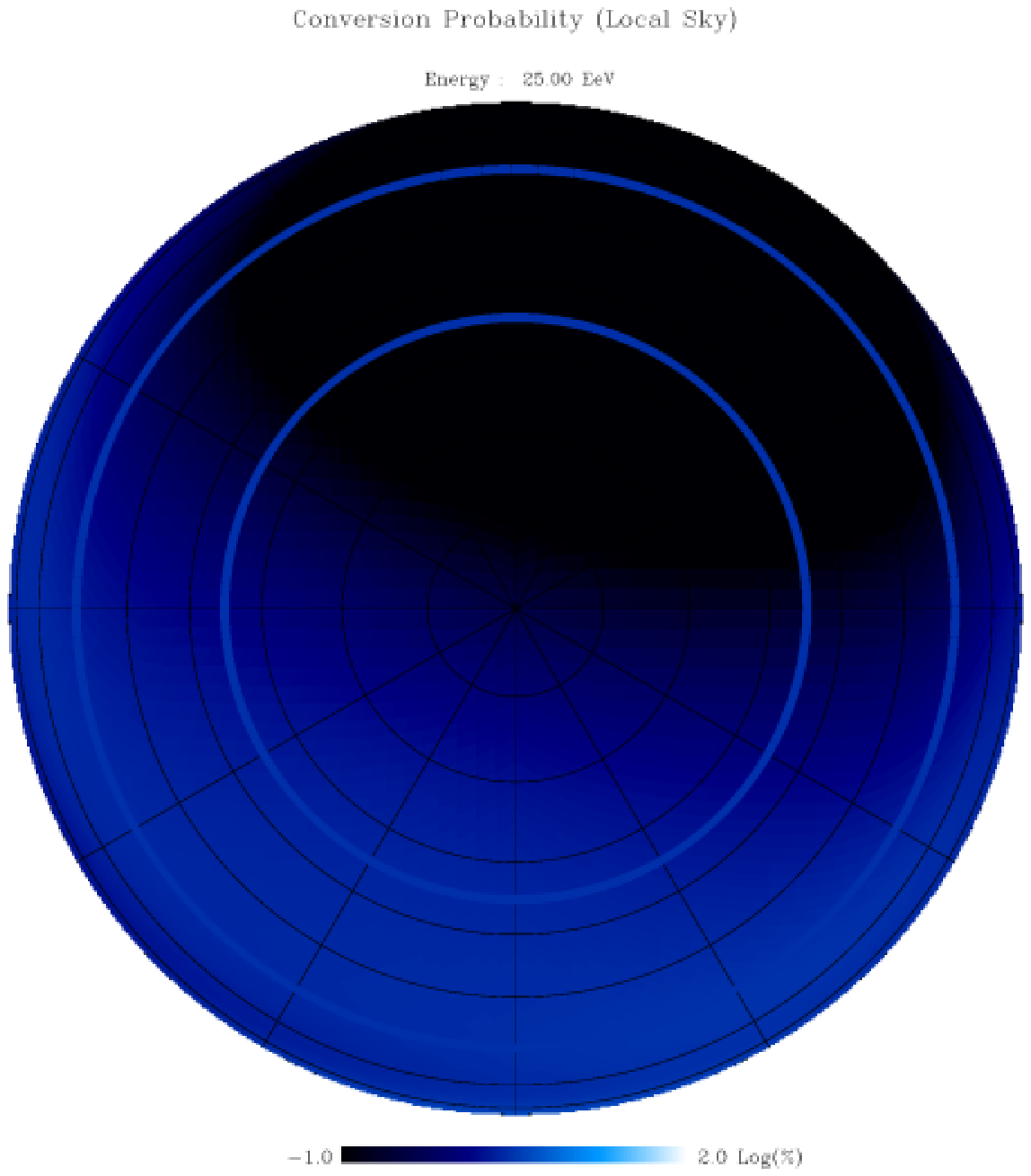}}}
\resizebox{10cm}{!}{\centering{\includegraphics{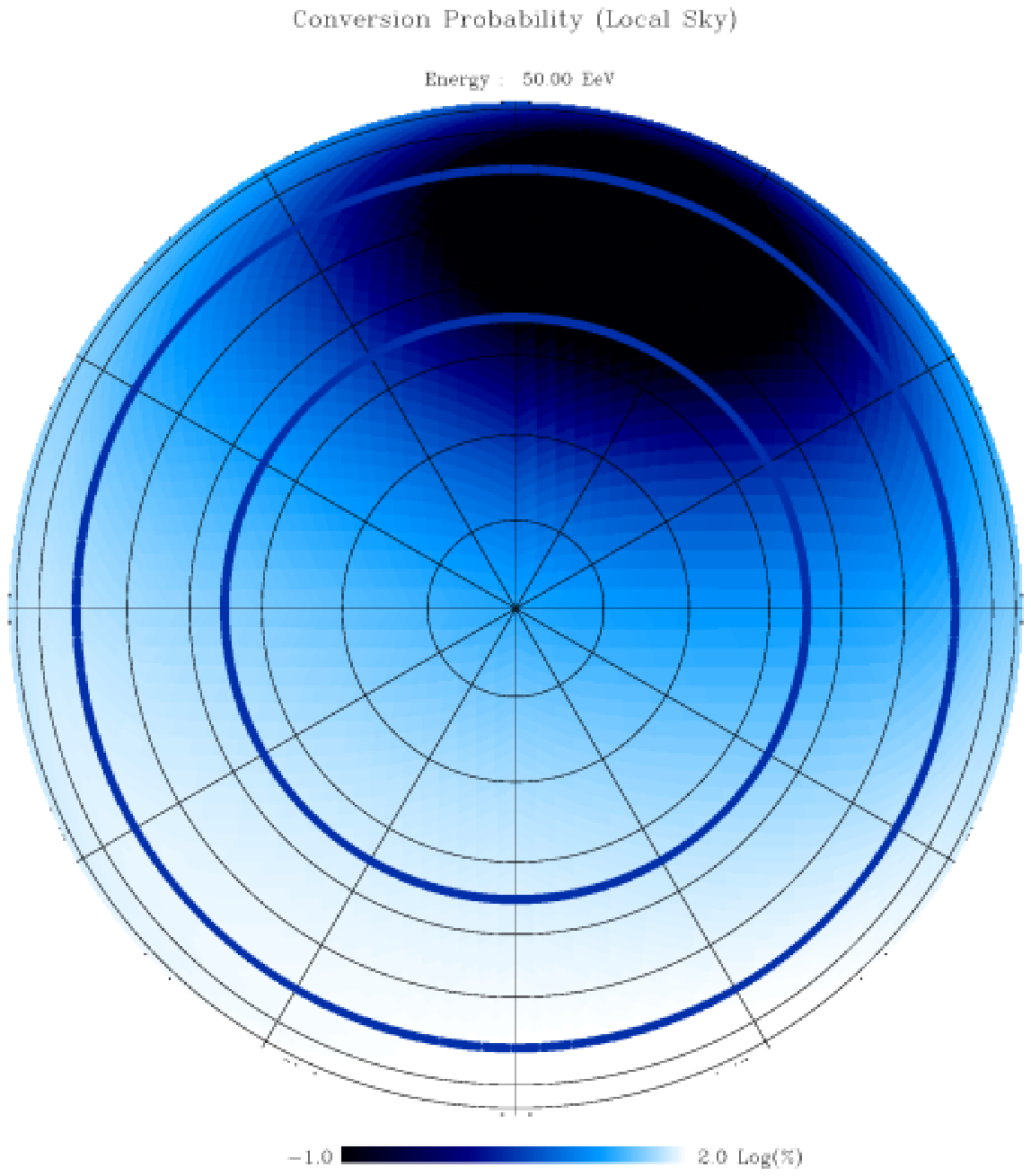}\includegraphics{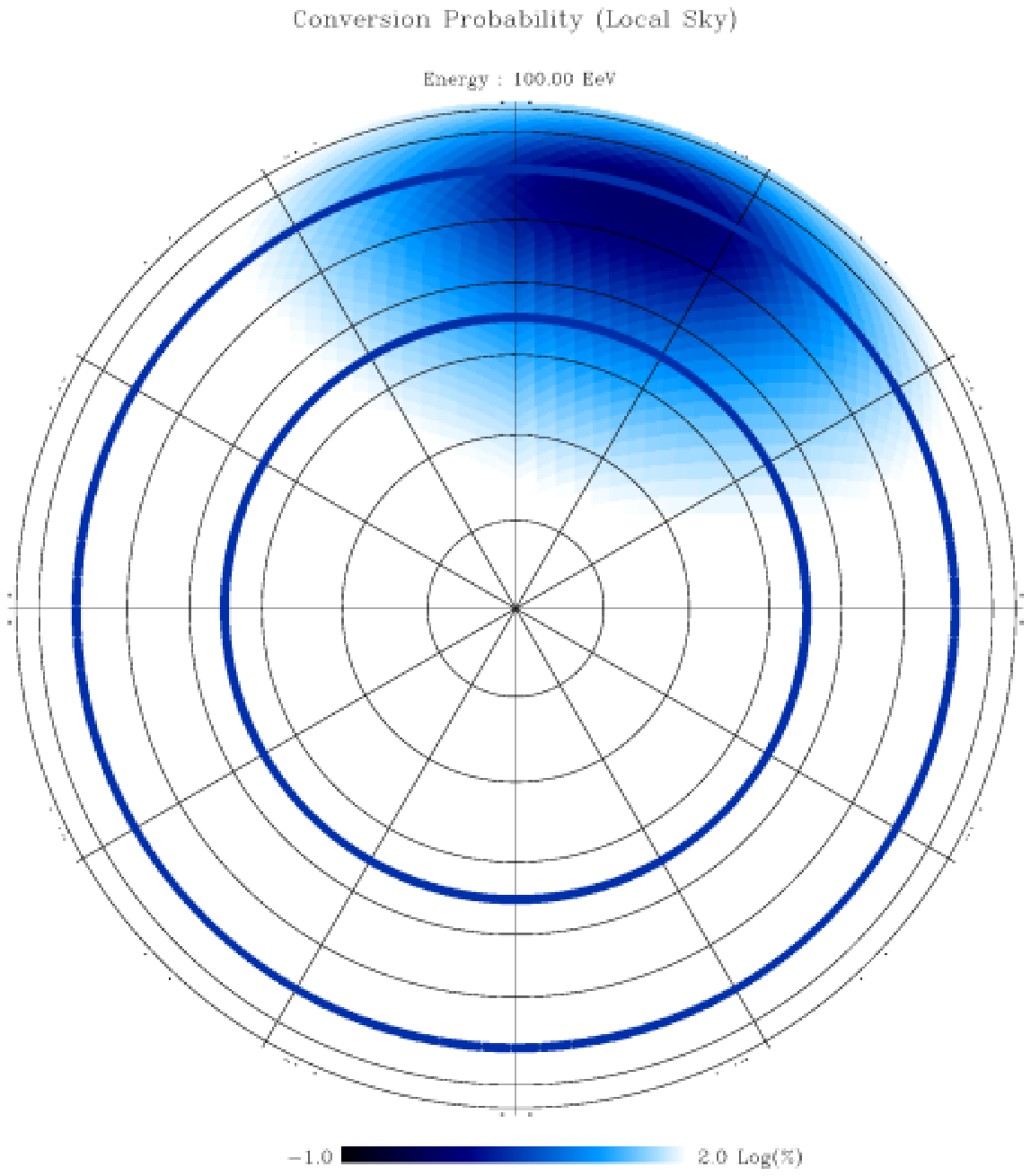}}}
\caption{\small{Maps of photon conversion probability for photons with
 $E_{prim}$ = 10EeV, 25EeV, 50EeV et 100EeV at Auger South (Latitude~: -35.22$^{\circ}$, 
 longitude~: -69.28$^{\circ}$).The zenith angles 35-60$^{\circ}$ indicated on theses maps with bold lines corresponds to the cuts applied on $\theta$ while performing the first analysis with SD described here. These maps have been made within an approximation of magnetic dipole for the geomagnetic field. }}
\label{maps}
\end{center}
\end{figure}

\begin{figure}[!h]
\begin{center}
\epsfig{file=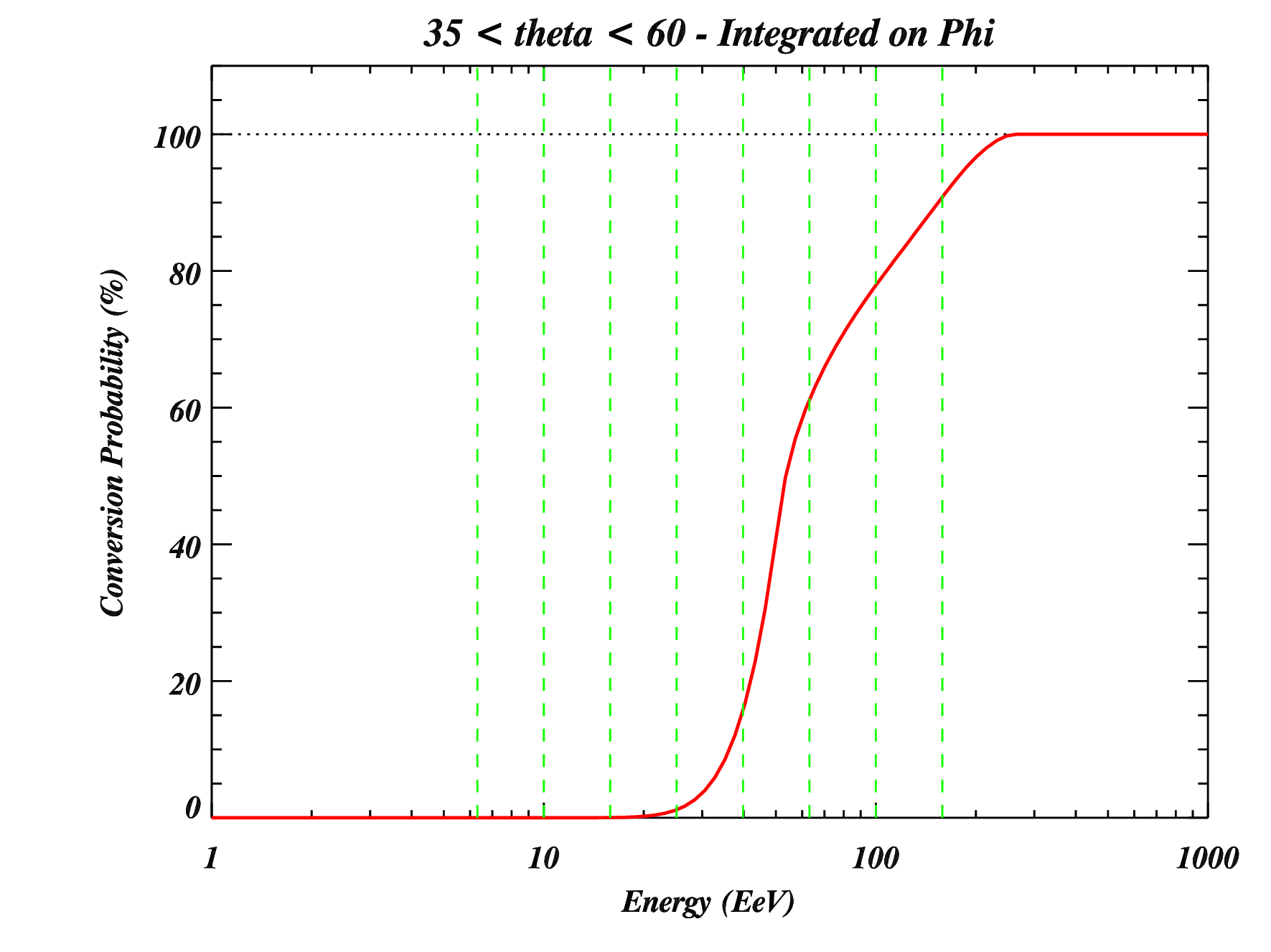, width=12cm}
\caption{\small{Conversion probability as a fonction of E integrated on all incoming directions between 35$^{\circ}$ and 60$^{\circ}$}}
\label{probaC}
\end{center}
\end{figure}

For the southern site of Auger, the conversion of the primary photon in a e$^+$e$^-$ pair as it enters the geomagnetic field becomes non negligible for energies above $5.10^{19}$ eV (Fig.~\ref{probaC}). This phenomenon results in an electromagnetic preshower entering the atmosphere. As a consequence, the particles of the preshower carry individually less energy than the initial photon, and the shower in the atmosphere is less affected by the LPM effect in case of conversion than in case of non conversion. This results in an earlier development of the photon shower (actually the $X_{max}$ of a converted shower never exceeds largely 1000g/cm$^2$ whereas the non converted showers have a typical $X_{max}$ above this value). \\

The conversion probability depends on the energy and $B_{\bot}$, the projection of the geomagnetic field perpendicular to the incoming photon direction (see \cite{MacBreen}). These considerations lead to the conversion probability maps shown on the figure Fig.~\ref{maps}. These maps show clearly that the conversion process has to be considered for $E_{prim} > $ 50 EeV. \\

As the development of a converted photon shower is earlier compared to that of an unconverted one (with the same characteristics), this additional effect has an impact on the method proposed herein :  
for converted showers, $X_{max}$ does not depend on $E_{prim}$ in the same way than for unconverted photons. If we don't know if the photon is converted or not,  the method described cannot be applied as we don't know which relation should be applied. Using some
average dependence between converted and unconverted ones to give the relation $S(1000)/ E_{prim}$ as a function of $\Delta X$ (e.g. using a combination of the two solutions weighted by
the probability of converting or not, accounting for the direction and approximate the energy of the event) can be thought of. Nevertheless, this is not a
satisfactory solution: because of the presence of a maximum in the distributions, the average
between two identical profiles at different positions, like the ones shown on fig.~\ref{univ} is not a profile at some
intermediate position. Even when correcting for such a bias, the resulting resolution in energy
would be poor. We have nevertheless several possibilities to deal with this issue and perform studies toward the setting of upper bound on the flux of photons. \\

A possible way out is to define, for each energy, regions in the sky where the conversion
is, either negligible, or almost sure. 
As a matter of facts, when applying the energy algorithms with both $X_{max}$ dependencies to candidate events, one can face different
situations:
\begin{enumerate}
\item the event is consistent with a surely converted photon, that is: when
applying the algorithm with the conversion hypothesis, the direction is within
the region of almost sure conversion at the estimated energy
\item the event is consistent with a surely unconverted photon (same
criterion).
\item the event is consistent with both.
\end{enumerate}
For the cases 1 and 2 we can build an analysis with a firm hypothesis on the conversion and choose correctly the energy converter we have to apply and the simulations we have to refer to for photon characterization purposes. As a consequence, one can build a sofisticated analysis that differs on the different regions of the sky. These regions change with the energy range considered  
(see Fig.~\ref{maps}, and pick up the lightest and the darkest regions for each energies, corresponding to case 1 and case 2 respectively). A conservative evaluation consists in counting cases 1 to 3 as candidates (twice for case 3, at two different energies), accounting for the acceptance in the angular regions defined above, as a function of energy.\\

One can stretch in addition that this scheme of analysis would bring another major improvement on the point of the energy resolution for the highest energy events~: the LPM effect lowers the photon/air cross-section which results in the already mentionned average delay in the first interactions but also obviously in fluctuations on the development. These fluctuations are becoming larger as the energy grows and the relation between $S(1000)/E_{prim}$ and $X_{max}$ becomes poorly defined at highest energies for unconverted photons. This could lead to serious problems  with the iterative procedure proposed here for the energy reconstruction. The analysis pattern proposed here would thus have an extra advantage~: with a single analysis, we rely more and more on converted photons as the energy grows. As converted photon showers suffer much less fluctuations as they are less affected by the LPM effect, the use of an energy converter depending on the development stage remain secure and accurate even at highest energies.

To perform a first analysis, we can even think of a simpler scheme than the one proposed just above. With 2 years of Auger data taking, one has to consider that the statistics of events available at the energies relevant for the conversion occurence is very small, especially if we cut the most vertical showers that would be seen well before their maximum and could lead to misreconstructions (indicated on Fig.~\ref{maps} by the inner circle). 
We could for a first step neglect the conversion, as we are mainly using events below the relevant energies for this phenomenon to happend. We can make the hypothesis that the potential photon candidates we might find would be all unconverted, and reconstruct them with the method and function described in section~\ref{univ}. Under this hypothesis, if there were converted photons, they would be obviously misreconstructed. Nevertheless, a conservative approach can be adopted to correct for this fact if we consider we have "lost" {\it all} the converted photons for the analysis when taking the upper bound on the photon flux. 
In the energy range  $10^{18.5}$ eV-$10^{19.2}$ eV,  were we have already large enough statistics to perform a relevant analysis, this assumption will lead to correct and conservative results.  Note that even if we set integrated upper limits, the corrections made on the upper bound will be very small in this energy range, due to the expected steepness of the  spectrum of cosmic rays photons and anyway leading to a conservative result. This solution is a simplistic approach, but can enable to set a robust first result on photon flux using the surface detector of Auger.\\

\section{Conclusion}

The reconstruction of the energy of photon induced showers with a surface detector can't be perform with a simple power law, as usually done for showers of proton or nuclear origin because of their late development. Parametrizations of the signals used
for showers initiated by a proton or nay other nucleus (which assume a factorization of the dependences on energy and zenith angle) are no longer valid. 
The explicit use of the development stage in the energy determination enables to build an efficient reconstruction method that leads to a similar resolution in energy for photon than the one we can currently expect for proton showers on Auger. An estimation of the $X_{max}$ of the shower has nevetheless to be done as an input to this method.



With a first guess for the energy of the primary, a precise evaluation may be
obtained from a relation between the interpolated signal at a given distance
from the core (e.g. $S(1000)$) and the stage of development expressed through
$X-X_{max}$ (the ``universal profile''). The energy of the primary is then reconstructed iteratively.
Note that, to some extent, the value of $X_{max}$
could even be inferred from
ground observables (e.g. time shape of the signals, curvature of the front) \cite{2004-010},
but these observables are also used to discriminate photon candidates from
30 showers, then it would be delicate to disentangle the discrimination
from the energy estimation. We have proposed here an iterative algorithm based
on the universal profile and on the averaged dependence of $X_{max}$ on
$E_{prim}$ for unconverted photons, which improves the resolution.\\

At energies above the magnetic conversion threshold, the mixing of
unconverted and converted photons makes the situation more complex.   
A conservative upper bound may be obtained by defining angular regions
(depending on energy) where the probability of conversion is close to either 0
or 1, and to evaluate the acceptance accordingly; of course, a 
{\em measurement} of the photon fraction as a function of the energy would be
more delicate. The current set of Auger data has still very poor statistics at energies where conversion occurs, so we can propose a simpler temporary analysis that relies on the unconverted photons and choose to loose efficiency as we would be unable to reconstruct converted photons. Here again the acceptance will have to be computed accordingly. This first analysis will obviously lead to a conservative result. One has nevertheless to keep in mind that all these algorithms (as well as analysis for the discrimination of
photons in itself) relies on computations of electromagnetic processes at ultra high energies and implicitely assume that QED may be used safely at these energies. 

\bibliographystyle{plain}
\bibliography{PetiteBiblio}


\end{document}